\documentclass[lettersize,journal]{IEEEtran}
\usepackage{graphicx}
\usepackage{array}
\usepackage{url}
\usepackage{hyperref}
\usepackage{xspace}
\usepackage{pifont}
\usepackage{xcolor}
\usepackage{multirow,makecell}
\usepackage{booktabs}
\usepackage{hyperref}
\usepackage{amsmath}
\usepackage[normalem]{ulem}
\usepackage{wrapfig,lipsum,booktabs}
\usepackage{xcolor}
\usepackage{soul}

\usepackage{pgfplots} 
\usetikzlibrary{patterns}

\usepgfplotslibrary{statistics}
\usepgfplotslibrary[statistics] 
\usetikzlibrary{pgfplots.statistics} 
\usetikzlibrary[pgfplots.statistics] 

\usepackage{esvect}
\usepackage{xspace}
\newcommand{\BfPara}[1]{\vspace{1mm}{\noindent\bf#1.}\xspace\xspace\xspace}
\hypersetup{
	plainpages=false,
	colorlinks,
	urlcolor=blue!50!black,
	linkcolor=red!50!black,
	citecolor=green!50!black,
	bookmarksnumbered
}

\usepackage{makecell}
\usepackage{array}
\newcolumntype{L}[1]{>{\raggedright\let\newline\\\arraybackslash\hspace{0pt}}m{#1}}
\newcolumntype{C}[1]{>{\centering\let\newline\\\arraybackslash\hspace{0pt}}m{#1}}
\newcolumntype{R}[1]{>{\raggedleft\let\newline\\\arraybackslash\hspace{0pt}}m{#1}}

\usepackage{multicol}

\newcommand{\etal}{{\em et al.}\xspace}

\usepackage{xspace}
\newcommand{\ours}{{\em Zad}\xspace}

\begin{document}
\title{Enhancing Vulnerability Reports with Automated and Augmented Description Summarization}

\author{Hattan Althebeiti, Mohammed Alkinoon, Manar Mohaisen, Saeed Salem, DaeHun Nyang, David Mohaisen 
\thanks{H. Althebeiti, M. Alkinoon, and D. Mohaisen are with the Department of Computer Science at the University of Central Florida (email: \href{mailto:mohaisen@ucf.edu}{mohaisen@ucf.edu}). M. Mohaisen is with the Department of Computer Science at Northeastern Illinois University, S. Salem with the Department of Computer Science at Qatar University and D. Nyang with Ewha Womans University. D. Mohaisen is the corresponding author (Phone: +1-407-823-1294).}
\thanks{A preliminary version of this work appeared in WISA 2022~\cite{abs-2210-01260}. This work was partially supported by the National Research Foundation (NRF) under grant 2016K1A1A2912757, the NVIDIA GPU Grant Program, and a seed grant from UCF’s Office of Research and Commercialization. H. Althebeiti also acknowledges funding from the Deanship of Graduate Studies and Scientific Research at Taif University.}
}

\maketitle

\begin{abstract}
Public vulnerability databases, such as the National Vulnerability Database (NVD), document vulnerabilities and facilitate threat information sharing. However, they often suffer from short descriptions and outdated or insufficient information.   In this paper, we introduce \ours, a system designed to enrich NVD vulnerability descriptions by leveraging external resources. \ours consists of two pipelines: one collects and filters supplementary data using two encoders to build a detailed dataset, while the other fine-tunes a pre-trained model on this dataset to generate enriched descriptions. By addressing brevity and improving content quality, \ours produces more comprehensive and cohesive vulnerability descriptions.  We evaluate \ours using standard summarization metrics and human assessments, demonstrating its effectiveness in enhancing vulnerability information.
\end{abstract}


\section{Introduction}\label{sec:introduction}

\IEEEPARstart{T}{he}critical need for shared vulnerability information led to the development of two key resources: the Common Vulnerabilities and Exposures (CVE) managed by MITRE~\cite{web_ref4}, and the National Vulnerability Database (NVD) managed by NIST~\cite{web_ref5}. CVE is a centralized repository that provides unique identifiers and essential attributes for newly discovered vulnerabilities. Security analysts and practitioners can leverage CVE to report and track vulnerabilities, ensuring clear and consistent communication across various industries. While largely synchronized, the NVD builds upon the CVE by offering in-depth analysis and incorporating additional information to help users prioritize and address vulnerabilities effectively.

CVE descriptions provide crucial context about vulnerabilities, e.g., affected software, versions, bug types, and potential impacts, and are pivotal for developers and security analysts to grasp the vulnerability's context, allowing a better understanding of how to develop the appropriate countermeasure. However, these descriptions can be brief, lack context, be inconsistent, outdated, or even incorrect ~\cite{nguyen2013reliability}.

Vulnerability descriptions are created by security analysts, and the written description hinges on their understanding of the vulnerability's attributes. This subjectivity is crucial, leading to gaps and inconsistencies in the disclosed descriptions and hindering their usability. Furthermore, enhancing and enriching these descriptions using available technical resources (e.g., hyperlinks to other reports) is complicated and challenging.

This paper aims to systematically enrich the existing vulnerability descriptions in NVD by adding supplementary materials to properly contextualize the original content and integrate information that was initially absent. This effort faces several challenges: first, identifying and selectively incorporating new and pertinent content requires a sophisticated resource discovery and summarization methodology. Secondly, given the diversity in the length and structure of potential resources, standardizing the presentation of these descriptions is necessary. Lastly, to ensure the enhancements meaningfully improve understanding and context, robust evaluation methods are essential to verify the achievement of these goals.

To tackle the challenges, we introduce \ours{}, a system that leverages an NLP pipeline to expand NVD vulnerability descriptions. This involves normalizing augmented descriptions against original entries, using a similarity metric to identify relevant resources. We frame the normalization as a summarization task, where the extended summary is provided as input alongside a cue for semantic relevance, and a pre-trained language model is fine-tuned for this specific task. The fine-tuning employs two strategies: label-guided, manually generated summaries for guidance, and summary-guided, utilizing the original summary as a semantic anchor. These approaches streamline the model's ability to efficiently produce relevant and contextually enriched summaries. Moreover, we refine our dataset using word frequency analysis to optimize content and minimize irrelevant details. For evaluation, we develop various human metrics to evaluate the generated text quality, which cannot be measured using conventional metrics to demonstrate the effectiveness of our approaches.

\BfPara{Contributions} (1) We present a pipeline that enriches the description of vulnerabilities by considering semantically similar contents from third-party resources. (2) To normalize the enriched description, we build \ours, an NLP pipeline that exploits advances in representation and pre-trained large language models that are fine-tuned using data in our problem domain. We fine-tuned the model on two distinct guides: the original short description and manually created summaries from the augmented text as labels to guide the model to generate normalized descriptions extractively. (3) We evaluate the performance of the \ours on NVD, a popular vulnerability database, with computational and human metric evaluations.

\section{Related Work}\label{sec:related}
NLP techniques have been applied in various domains, including security repository analysis~\cite{shahid2021cvss,ChenKMAA20}, privacy-policy classification~\cite{alabduljabbar2021tldr}, and malware detection~\cite{mohaisen2015amal}. Alabduljabbar \etal~\cite{alabduljabbar2021tldr} classified textual segments in third-party privacy policies, assigning high-level labels of distinct data privacy practices. They conducted extensive experiments on different embeddings and learning algorithms, selecting the best-performing model to build an ensemble classifier. Their approach achieved over 90\% accuracy in segment labeling.

Shahid \etal~\cite{shahid2021cvss} used Bidirectional Encoder Representations from Transformers (BERT)~\cite{BERT} to estimate the characteristics and severity of vulnerabilities using the Common Vulnerability Scoring System (CVSS) framework.
The CVSS score uses multiple metrics, each labeled based on severity and effects.
The authors deploy multiple instances of BERT and use the vulnerability description to predict the labels for each metric. Each metric is used to compute the severity score based on the CVSS formula.

Prior studies have applied NLP to analyze vulnerability databases, addressing their limitations~\cite{wu2021literature,AnwarKNM18,AnwarACLM22,anwar2020measuring} and extracting industry-specific information~\cite{xu2019analysis}. Dong \etal~\cite{ref_proc1} developed a system to detect inconsistencies between NVD/CVE and third-party reports using a Named Entity Recognition (NER) and a Relational Extractor (RE). 

Guo \etal~\cite{guo2020predicting} addressed the incompleteness issue in NVD and utilized private repositories, namely, IBM-X-Force~\cite{ibm} and SecurityFocus~\cite{bugtraq}, to supplement the original description.  Kühn \etal~\cite{kuehn2021ovana} developed OVANA, a system that enhances the vulnerability's description in the NVD and measures the Information Quality (IQ) using a set of metrics, namely, accuracy, completeness, and uniqueness.
OVANA utilizes NER to extract names from the description, which are used to predict the CVSS score for a vulnerability.

Mumtaz \etal~\cite{mumtaz2020learning} developed a method for learning vulnerability-focused word embeddings by compiling a dataset from multiple sources and training Word2Vec~\cite{word2vec}. The learned embeddings effectively captured relationships between security terms like viruses and malware. Similarly, Guo \etal~\cite{guo2020predicting} extracted features from CVEs to curate a dataset and trained a Word2Vec model on CVE descriptions to learn targeted embeddings. These embeddings were then used to train a Neural Network (NN) classifier to identify missing features in new vulnerability descriptions. Yitagesu \etal~\cite{yitagesu2021automatic} employed PenTreebank (PTB) to train an NN for tagging technical terms, creating an annotated corpus from vulnerability descriptions.

Researchers have explored the relationship between CVEs and other attributes, such as Common Attack Pattern Enumeration and Classification (CAPEC), which describes attack patterns used to exploit vulnerabilities. Kanakogi \etal~\cite{ref_article1,ref_proc3} employed three types of embeddings to encode CVE descriptions and CAPEC entries, using cosine similarity to assign the most relevant CAPEC to each CVE. Similarly, W{\aa}reus and Hell~\cite{ref_lncs1} developed an NLP-based method to automatically assign Common Platform Enumeration (CPE) to CVEs, where CPE identifies vulnerable software versions.

Malware remains a major threat, with malware detection and classification extensively studied using NLP~\cite{mohaisen2013unveiling,mohaisen2014av,mohaisen2015amal}. Incident reports, often unstructured, require normalization for effective threat intelligence. Upadhyay \etal~\cite{upadhyay2022mapping} developed an NLP pipeline that maps extracted information to the cyber kill chain model, aiding incident report normalization. Similarly, Niakanlahiji \etal~\cite{niakanlahiji2018natural} introduced SECCMiner, a system for extracting key concepts from APT reports.

\BfPara{Additional Applications} The primary outcome of \ours is enriched contextual vulnerability descriptions, enabling various security applications explored in the literature, such as exploitability analysis~\cite{YinTCW20,0003TC0YL22,YinTCYWA23}. A context-rich description enhances the precision and effectiveness of these applications.

\section{\ours: Building Blocks}\label{sec:method}
We first address challenges \ours alleviates (\textsection\ref{subsec:challenges}), then an overview of the pipeline (\textsection\ref{subsec:overview}). Next, we introduce the sequence encoder (\textsection\ref{sec:se}) and explain its role within its pipeline. Finally, we present two pre-trained models commonly used in our summarization (\textsection\ref{sec:pre-trained}).

\subsection{Challednges}\label{subsec:challenges}
\ours enhances vulnerability descriptions in public databases using third-party reports. However, systematically achieving this goal presents several challenges:

\BfPara{Challenge 1: Augmentation Technique}
\ours employs augmentation to enhance vulnerability descriptions, guided to ensure relevance. We frame this as a semantic search problem, retrieving relevant excerpts from external reports. During augmentation, cosine similarity measures the relevance and similarity between the original and added content.

\BfPara{Challenge 2: Text Encoding}
Encoders project data into a high-dimensional space, with text encoding occurring at word or sentence levels. Word-level encoding captures a word's contextual semantics, while sentence-level encoding preserves broader structural and semantic features. For semantic search, we use sentence-level encoding to effectively represent sentence nuances within the embedding space.

\BfPara{Challenge 3: Evaluation}
\ours generates summaries from augmented text while preserving the original description’s essence. Given the critical nature of vulnerability reports, accurate evaluation is essential. Traditional assessments rely on word overlap with reference summaries but fail to capture key details like software names or specific bugs. They also overlook linguistic aspects such as fluency and comprehension. To address this, we introduce human-centric metrics that assess fluency, completeness, correctness, and understanding for a more holistic evaluation.

\begin{figure*}[t]
    \centering
    \includegraphics[width=1\textwidth]{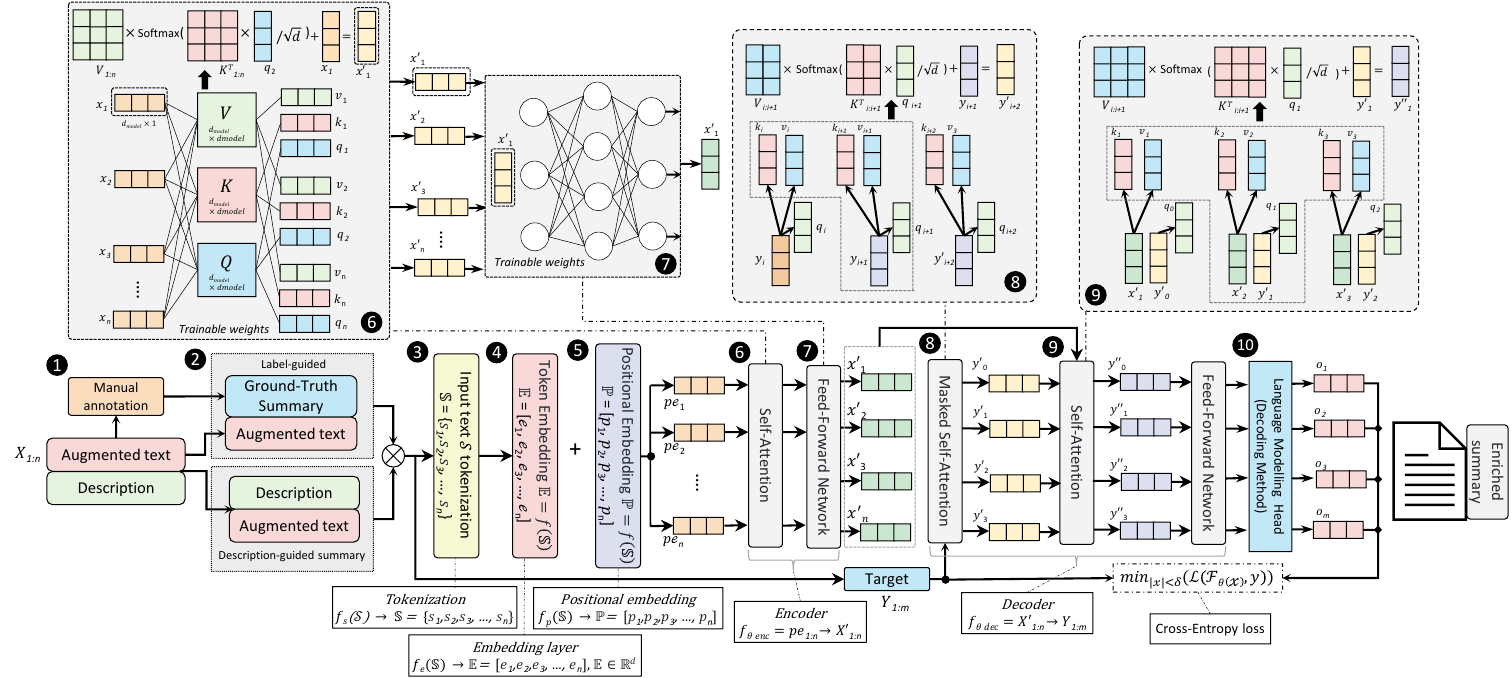}
    \caption{\ours pipeline. The pipeline consists of multiple steps to transform the text $X_{1:n}$ from a series of tokens into numerical representations. The encoder component then encodes them to incorporate their context, improving their embeddings through self-attention. The decoder's role is to deploy the encoded sequences to predict the target sequence $Y_{1:m}$.}
    \label{figure:Vul_Aug_pipeline}\vspace{-5mm}
\end{figure*}

\subsection{An Overview of \ours}\label{subsec:overview}
Now we present an overview of \ours's pipeline, as shown in Figure~\ref{figure:Vul_Aug_pipeline}, which includes dataset selection and the transformer model~\cite{transformer}. The dataset pairs augmented text from third-party reports as input with the vulnerability description as the target.  

In step \ding{182}, we introduce a label-guided approach by selecting 100 samples from the dataset and manually creating summaries based on the augmented text and description. Annotation details and datasets are discussed in Section~\ref{sec:dataset}.

Step \ding{183} presents two fine-tuning approaches, each using a separately deployed dataset. The Seq2Seq model~\cite{seq2seq} maps an input sequence of length $n$ to a target sequence of length $m$ via a function $F: X_{1:n} \rightarrow Y_{1:m}$.  After selecting the training dataset, the augmented text undergoes tokenization in \ding{184}, breaking it into tokens. In \ding{185}, tokens are projected into a low-dimensional space as embeddings of size $d$.

Transformer models use self-attention, which ignores word order. In step \ding{186}, positional embeddings are added to token embeddings to retain positional information. The final token representation is passed to the self-attention layer in step \ding{187}, where each token attends to all others. In step \ding{188}, tokens pass independently through a Feed-Forward Network (FFN) to refine their embeddings. Steps \ding{187} and \ding{188} form the transformer encoder, enriching the input representation.

The decoder generates the target \( Y_{1:m} \) from the encoded sequence. In step \ding{189}, the target text is masked with self-attention, restricting access to future tokens. The masked attention output and step \ding{188}'s output pass through another self-attention layer in step \ding{190}, followed by an FFN. In step \ding{191}, the output sequence is sent to the language modeling head to predict the next token. The model updates parameters by computing the loss between predicted and target tokens. During training, the target token is fed back autoregressively, generating tokens until the special end-of-sequence token \( <\text{EOS}> \) appears.

\ours enriches the description through the extended content of text while minimizing the semantic gaps with the target summaries.
The objective is to find a function $\mathcal{F}$ parametrized by $\mathcal{\theta}$ to map text $x$ under length constraint $\delta$ to produce output $y$. More formally, the loss function is defined as:
\begin{equation}
\min _{|x|<\delta}\left(\mathcal{L}\left(\mathcal{F}_\theta(x, y)\right)\right.
\end{equation}
$\theta$ for the pre-trained model is adjusted according to the objective function and its number of parameters is dependent on the underlying pre-trained model. 

\subsection{Sentence Encoders}\label{sec:se}
We use a sentence encoder for its well-established benefits. To select the most suitable encoder, we evaluated multiple candidates on various CVEs and their associated third-party reports. Our results identified the Universal Sentence Encoder (USE)~\cite{USE} and MPNet~\cite{MPnet_se} as the best options.

\BfPara{Universal Sentence Encoder (USE)}  \label{USE}
USE, introduced in~\cite{USE}, offers two architectures with trade-offs. The transformer-based model provides accurate, context-aware representations with a complexity of \( O(n^2) \), where \( n \) is the sequence length. The Deep Averaging Network (DAN)~\cite{DAN} simplifies this by averaging word and bi-gram embeddings, reducing complexity to \( O(n) \). Given our dataset's variability, domain specificity, and scalability needs, we prioritize DAN’s efficiency over the transformer's accuracy.

\BfPara{MPNet} MPNet~\cite{MPNet} combines the strengths of BERT~\cite{BERT} and XLNET~\cite{XLNET}. BERT uses Masked Language Modeling (MLM), masking 15\% of tokens for prediction but ignores dependencies between them. XLNET, in contrast, employs Permuted Language Modeling (PLM), considering token permutations but causing position discrepancies between pretraining and fine-tuning. MPNet unifies both objectives, capturing token dependencies while preserving positional integrity. An MPNet sentence encoder is fine-tuned on diverse datasets to generate unique sequence representations.

\subsection{Pre-trained Models}\label{sec:pre-trained}
\ours fine-tunes pre-trained models on our datasets for augmentation. These models leverage transfer learning~\cite{krizhevsky2017imagenet,simonyan2014very,szegedy2015going}, where a neural network is first trained unsupervised on general tasks, such as predicting the next or masked token, and then fine-tuned on a specific dataset. Transfer learning for NLP was introduced in~\cite{ulmfit} and later adopted in transformer models, which retain the original transformer architecture~\cite{transformer}.

The transformer comprises two components: an encoder and a decoder. The encoder builds a representation for the input sequence to capture the dependencies between tokens in parallel without losing their positional information. 
The transformer relies on attention mechanisms to capture the interdependencies within a sequence, providing a context-aware representation for each token. 
The decoder uses the built representation and maps it to a probability distribution over the vocabulary to predict the next token.

The original transformer was designed for machine translation~\cite{wang2019learning} but later generalized to various tasks with exceptional results. Most modern pre-trained models adopt a transformer architecture using either an encoder (e.g., BERT~\cite{BERT}), a decoder (e.g., GPT~\cite{GPT}), or both. Each design excels in specific tasks. Summarization, for instance, is best modeled as a sequence-to-sequence task, where an encoder-decoder architecture is ideal for transforming text into concise summaries.

The leading models for summarization are BART~\cite{BART}, T5~\cite{T5}, and Pegasus~\cite{Pegasus}, with BART and T5 being the most widely used. BART, a denoising autoencoder for Seq2Seq pretraining, corrupts text with a noising function and trains the model to reconstruct the original text. In contrast, T5 employs MLM objectives like BERT but masks text spans as its corruption strategy, with span length affecting performance only when excessively long. T5 also introduces a task-specific prefix framework, enabling a single model to support multiple NLP tasks by learning task associations from training data.

\section{Pipeline and Technical Details}\label{sec:pipeline}

Now we discuss the steps outlined in the pipeline depicted in Figure~\ref{figure:Vul_Aug_pipeline}.
The pipeline illustrates a transformer architecture with an encoder and a decoder leveraged for our task. 

\BfPara{Annotation}
We manually annotated 100 samples from the collected dataset, creating a summary for each one from the augmented text.  
Then, two datasets are prepared for training: description-guided and label-guided datasets, as depicted in~\ding{183}. 
Next, we describe the details of each dataset. 

\BfPara{Description-guided dataset}
This dataset employs the description as a label for summarization, whereby the model is directed through the description to reduce the semantic discrepancies between the generated text and the description. This method relies on the model to produce concise and enhanced summaries. However, this approach limits the model to brief descriptions, which limits its capacity to create longer summaries. Moreover, the content of the augmented text for some vulnerabilities is very long compared to its corresponding description, making learning difficult. 

\BfPara{Label-guided dataset}
The augmented text relies on similarity measures, making it vulnerable to issues like highly similar but contextually different text. To address this, we randomly selected 100 samples and manually created ground-truth summaries.  Before summarization, we considered key factors. Given the sparse and unstructured nature of the augmented text, we adopted an extractive approach, ensuring conciseness and coherence. Version numbers were included only when minimal, prioritizing summary quality. In rare cases, we retained the original description when it contained crucial missing details. Since our summaries are extractive, they must accurately reflect the specific CVE. However, some augmented texts included details from other CVEs, requiring manual verification for accuracy.

\BfPara{Tokenization} Tokenization entails breaking a text into separate and independent entities called tokens. 
For instance, the tokenization could be based on a unigram, where each word is represented as a single token.
Tokenization is a mapping that maps a character string to a set of tokens. 
Given an alphabet $\Sigma$, the set of all possible character strings defined over the alphabet is $\Sigma ^*$, and the set of all tokens
is denoted by $D$.
Tokenization  is a mapping that takes a character string representing a text
\(S \in \Sigma ^*\) as an input and outputs a set of tokens, \(F_D(S)=\{a_1,\hspace{1mm}a_2,\hspace{1mm}a_3,\hspace{1mm}a_4, \dots,\hspace{1mm}a_n\}\), where $a_i \in D$.

Tokenization can occur at the word, sub-word, or character level. Word tokenization splits text using spaces or punctuation, adding each word to the vocabulary (vocab). However, this can lead to a large vocab size, increasing the embedding matrix's dimensionality and making training and inference costly. To mitigate this, vocab size is typically limited to the most common 100,000 words in the training corpus, with unknown words encoded as \( <\text{UNK}> \).

Character-level tokenization reduces vocabulary size and computational costs by splitting text into individual characters, making it effective for handling unknown words. However, it sacrifices linguistic structure and semantics. Sub-word tokenization balances word and character tokenization by breaking words into smaller units, preserving meaning while handling complex words and morphological variations. BART and T5 exemplify this approach, using Byte Pair Encoding (BPE)~\cite{ref_proc15} and SentencePiece~\cite{kudo2018subword}, respectively, to optimize vocabulary use and enhance text representation.

\BfPara{Byte Pair Encoding (BPE)} Initially developed for text compression~\cite{gage1994new}, BPE was later adapted to represent open vocabulary with a limited size using variable-length character sequences. Many pre-trained models, including GPT~\cite{GPT}, GPT-2~\cite{radford2019language}, and RoBERTa~\cite{liu2019roberta}, utilize BPE as a tokenizer.  BPE segments words into sub-units, enabling efficient representation of both common and rare words while reducing vocabulary size. Multiple words can share sub-units, improving generalization. For the rest of our discussion, we focus on its application in the English language.

BPE initializes a base vocabulary of ASCII and select Unicode characters. It expands by merging frequent adjacent character pairs into new tokens while adhering to a predefined vocabulary size limit. This process continues iteratively until the limit is reached, optimizing tokenization efficiency.

\begin{figure}[t]
    \centering
    \includegraphics[width=0.43\textwidth]{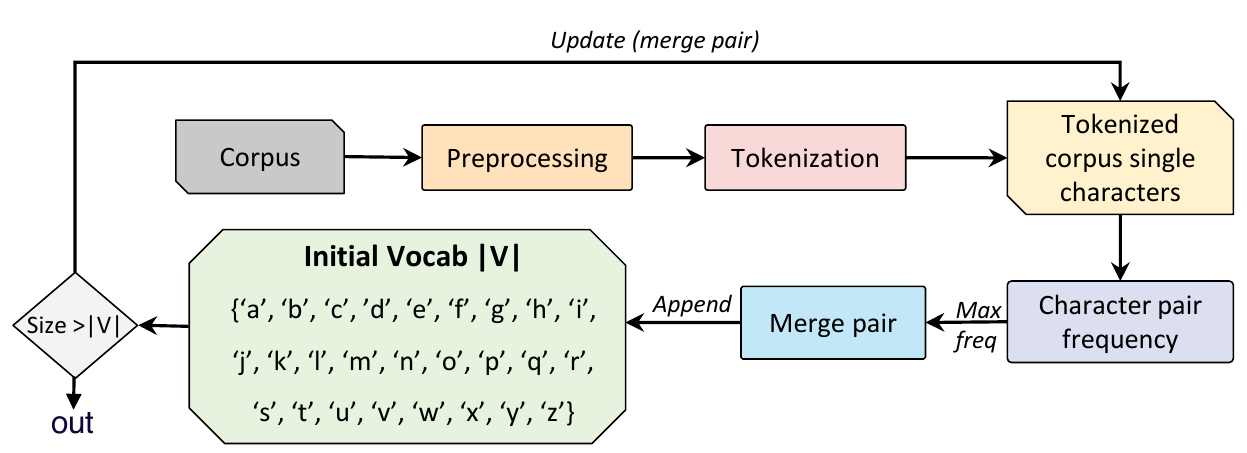}
    \caption{A schematic structure for training Byte-Pair Encoding (BPE) tokenizer}
    \label{figure:BPE}\vspace{-5mm}
\end{figure}

The tokenizer is trained on a large raw text corpus representative of its application domain. It includes a special token, such as \( <\text{UNK}> \), for out-of-vocabulary words not learned during training or exceeding the vocabulary size limit. When applied, the tokenizer segments text into known tokens, mapping unknown characters or tokens to \( <\text{UNK}> \). Figure \ref{figure:BPE} illustrates the BPE tokenizer pipeline and its procedural steps.

\BfPara{SentencePiece} SentencePiece is another sub-word segmentation algorithm based on the Unigram language model. 
The Unigram model is a statistical model with two assumptions. (1) It assumes the occurrence of each word is independent of its previous words. 
(2) A word's occurrences depend on its frequency in the dataset. The probability of a sequence $\mathbf{x}$ , represented as $\mathbf{x}=\left(x_1, \ldots, x_m\right)$, can be modeled as:
\begin{equation}
P(\mathrm{x})=\prod_{i=1}^m p\left(x_i\right), \forall i, \hspace{1mm} x_i \in \mathcal{V},
\end{equation}
where $\mathcal{V}$ is the vocabulary set and $m$ is the number of tokens in the sequence, e.g., $P(x_1, x_2, \dots, x_m)= P(x_1) \times P(x_2) \times \hdots \times P(x_m)$. 
SentencePiece starts with an extensive vocabulary set obtained from a corpus and works by removing tokens from that set until it reaches the desired size. All the words presented in the corpus and their most frequent sub-strings, along with all individual characters, are used as the base vocabulary for the model~\cite{kudo2018subword}.

SentencePiece utilizes a unigram model for vocabulary optimization by identifying and removing the least effective tokens. It calculates the probability of token combinations to find the most effective tokenization for words, aiming to achieve the highest score. The model assesses the impact of token removal by computing the loss, which considers word frequency and probability. Vocabulary optimization is achieved by iteratively removing tokens and reassessing the loss.

\BfPara{Positional Embedding}
Positional embeddings preserve token order by disrupting self-attention's permutation equivariance, ensuring shuffled tokens do not yield identical representations. An effective embedding method must uniquely encode each position, maintain a constant distance between consecutive tokens, and generalize to longer sequences. Positional embeddings can be either learned during training or predefined using a fixed function.

\BfPara{\em Positional Embedding (PE)} PE, used in BART, assigns an embedding vector for each position, defining a unique embedding for each token in the sequence. Therefore, the same word will have different embedding at different positions. One advantage of this method is its simplicity and effectiveness. In contrast, PE cannot encode sequences longer than the longest sequence observed during training, which is an issue if the maximum length is unknown. 

\BfPara{\em Relative Positional Encoding (RPE)} RPE employs a dynamic method to learn a word's position using the attention layer within the encoder.
We focus on the role of the attention layer in integrating positional information.
The attention layer computes how much attention a word should pay to other words within a sequence.
RPE uses positional embedding by analyzing the pairwise relationship between words within a sequence. 
The input representation is modeled as a fully connected directed graph. 
For tokens $x_i$ and $x_j$ within a sequence, two edges are represented as $a_{i j}^V$ and $a_{i j}^K$. 
Edges' representation is added to the attention layer to determine a token's position relative to others, prioritizing relative over absolute position and enhancing token relationships' encoding.

\BfPara{Transformer} The transformer is the foundation for all large language models composed of an encoder and a decoder. The encoder uses self-attention and eliminates the recurrence in RNN.  Self-attention is parallelizable and can compute the attention of individual tokens within a sequence without the need for memory resources or any sequential processing.

\BfPara{\em Encoder} 
The encoder is composed of two sub-layers: the self-attention layer in step \ding{187} and the Feed-Forward Network (FFN) in step \ding{188}.
The self-attention layer deploys attention pooling, an operation capable of aggregating inputs to produce an output.
There are various types of attention-pooling functions. 
However, the most common one is scaled dot-product attention, which is defined as:  
\begin{equation} \label{equ:attention}
\operatorname{Attention}(Q, K, V)=\operatorname{softmax}\left(\frac{Q K^T}{\sqrt{d}}\right) V.
\end{equation}

The self-attention layer in \ding{187} accepts a sequence, \(x = (x_{1}, x_{2},\dots, x_{n})\) of length $n$, in which \(x_{i} \in {R}^d\), and produces a new sequence \(z = (z_{1}, z_{2},\dots, z_{n})\) such that \(z_{i} \in {R}^d\) captures how much attention $z_i$ should have with respect to the sequence \((x_{1},\dots, x_{n})\).
The transformer improves self-attention using query, key, and value.

Self-attention applies a weighted sum to each token in a sequence to encode its relationships with others. As shown in (\ref{equ:attention}), each token is projected into query \( Q \), key \( K \), and value \( V \) representations through linear layers. The query of a token is compared to the keys of all tokens in a sequence of length \( n \) using the dot product. However, as the embedding size increases, the dot product grows, potentially affecting the gradient. To mitigate this, dividing by \( \sqrt{d} \), where \( d \) is the embedding size, stabilizes gradients and maintains weight scaling. The resulting vectors pass through a softmax function, producing a probability distribution, which is then multiplied (dot product) by each token's value vector and summed to form the final embedding, encoding dependencies of the sequence.

Multi-headed attention consists of multiple self-attention layers, or ``heads,'' capturing diverse semantic and syntactic characteristics of a sequence. Each head focuses on different token relationships, improving representation. The outputs from all heads are concatenated and passed through a linear projection layer to generate the final sequence representation.

The Feed-Forward Network (FFN) in \ding{188} is a fully connected layer without activation, enhancing the model's expressive power. As shown in~\cite{FFN_mem}, FFN functions as key-value memory, where key vectors encode sequence patterns and value vectors represent token distributions. Each token passes through FFN independently to refine its embedding. Self-attention accounts for one-third of trainable parameters, while FFN consumes the majority. The original transformer utilized six encoder blocks in a cascade, where each encoder's output serves as input to the next, with the final encoder output passed to the decoder.

\BfPara{\em Decoder} The decoder leverages the hidden states obtained from the encoder and previously generated tokens to predict the next token at each time step $t$.
The decoder has two distinctions from the encoder.
The masked self-attention in \ding{189} prevents the decoder from attending to future tokens in the target sequence during training by allowing access to the target sequence to evaluate its prediction.

The self-attention layer operates similarly in \ding{187} but with a key difference: the query comes from the masked self-attention layer output, while key and value vectors are from the encoded sequence. The decoder's final output feeds into the LM head, which consists of a linear layer followed by softmax to generate a probability distribution over the vocabulary $\mathcal{V}$. Additionally, the LM head specifies a decoding method to select the next token based on this distribution. 
After selecting a token as its prediction, the predicted token is used to compute the cross-entropy loss with the target token for that time step and update its parameters until it reaches the $<$end$>$ token. 

Summarization relies on autoregressive text generation, predicting the next token based on previous ones. This process factorizes conditional probabilities using the chain rule but can suffer from numerical instability. To address this, logarithms of conditional probabilities are used, converting the product into a sum.  Decoding follows two key principles:  (1) Iterative Selection: The next token is chosen based on the sequence generated up to the current time step \( t \). (2) Task-Specific Emphasis: Decoding prioritizes different characteristics; for summarization, sequence quality is key, whereas in open-domain conversation, diversity is more critical.

\BfPara{Decoding} We use beam search for decoding, as summarization prioritizes factual accuracy. This method, parameterized by the number of beams, maintains multiple hypotheses (sequences) during generation. It extends partial hypotheses by appending probable tokens until the end-of-sequence token is reached. Sequences are then ranked by log probabilities, selecting the highest-scoring one.  Alternative methods include top-k sampling, where a parameter \( k \) limits token selection to the top \( k \) highest-probability tokens, with one chosen randomly. Nucleus sampling introduces a probability threshold \( p \), selecting a dynamic subset of tokens whose cumulative probability exceeds \( p \), redistributing probability within this subset to allow adaptive token selection.

\BfPara{Operational Considerations} Transformers are commonly deployed in two settings:
 (1) Feature extraction involves freezing model parameters to compute hidden states for each word embedding. Only the classification head is trained on task-specific data, which is fast and resource-efficient.
    (2) Fine-tuning entails adjusting all trainable parameters for the task at hand. This process demands time and computational resources, particularly for larger models.
    In our scenario, we fine-tune BART and T5 models. Since BART has fewer parameters, its fine-tuning (both training and testing) is faster.

\section{Evaluation}\label{sec:eval}
\subsection{Dataset and Data Augmentation}\label{sec:dataset}
\subsubsection{Data Source and Scraping} Our data source is NVD, which comes with many strengths, including (1) detailed structured information, including severity score and publication date, (2) human-readable descriptions, (3) capabilities for reanalysis with updated information, and (4) powerful API for vulnerability information retrieval.  In our data collection, we limit our time frame to vulnerabilities reported between 2019 and 2021 (inclusive). Our analysis shows CVEs reported before 2019 do not include sufficient hyperlinks with helpful content, which is our primary source for augmentation.  We scrape the links for all vulnerabilities in our time frame, totaling $35,657$ vulnerabilities. For each vulnerability, we scrape the description and the associated hyperlinks, which will be scraped next for augmentation. 

\subsubsection{Description Augmentation}
First, we iterate through the scraped hyperlinks, which direct to third-party pages, either official (vendor/developer) or unofficial (e.g., GitHub issue tracking). For each page, we extract text from paragraph tags (\(<\textit{p}>\) \(<\backslash\textit{p}>\)) separately and apply preprocessing to clean the text. This includes removing web links, special characters, redundant white spaces, phone numbers, and email addresses.  Next, we check paragraph length, discarding those shorter than 20 words, as they likely lack sufficient details for our purpose.

Next, we encode the semantics of the extracted paragraphs and scraped descriptions into low-dimensional vector representations, as established in Section~\ref{sec:se}. We compute cosine similarity between these vectorized representations, yielding a value between \(-1\) and \(1\).  Let \(\mathbf{v}_p\) and \(\mathbf{v}_d\) represent the vectorized paragraph and description, respectively. Cosine similarity is defined as:  \(
\cos(\mathbf{v}_p, \mathbf{v}_d) = \frac{ \vv{v_p}\cdot \vv{v_d} }{||\vv{v_p}||\hspace{0.1cm}||\vv{v_d}||}
\). If \(\cos(\mathbf{v}_p, \mathbf{v}_d)\) exceeds a threshold, the paragraph is added to the augmented text. This process is repeated for all paragraphs on a page and across all hyperlinks. After processing, we create an entry with the augmented text as input and the description as the target. Vulnerabilities without hyperlinks or with no paragraphs meeting the threshold are excluded from the dataset.

\begin{figure}[t]
    \centering
    \includegraphics[width=0.48\textwidth]{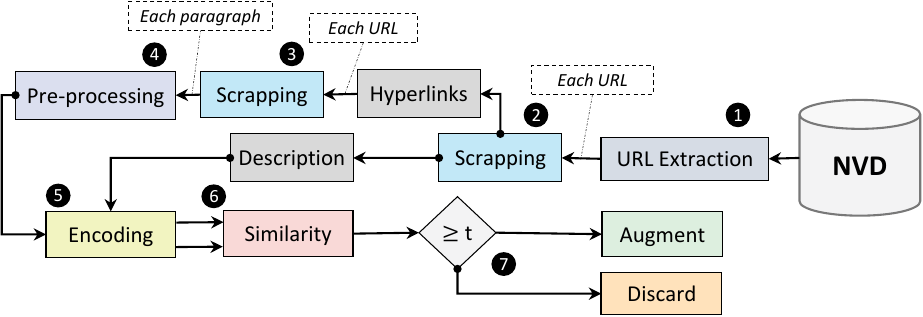}
    \caption{Our data collection pipeline.}
    \label{figure:Data_collection_pipeline}
\end{figure}
 
Figure~\ref{figure:Data_collection_pipeline} shows the pipeline for our data collection.
The choice of a sentence encoder will affect the dataset.
Our preliminary assessment found that USE is more accurate (sensitive) than MPNet in terms of the similarity score representation---when the description and the paragraph are (semantically) similar to one another, USE produces a higher score than MPNet and vice versa. 

We set the similarity score for USE to be between $0.60$ and $0.90$. Since MPNet is less sensitive, we use a more restrictive threshold between $0.70$ and $0.90$. 
We excluded paragraphs with a similarity score above $0.90$ because we found them to be exact to the description, and adding them would not enrich the description. 
Those values were picked as part of our assessment of the two encoders using a small set of vulnerabilities and following the above procedure. Therefore, our pipeline builds two datasets using each encoder. 

We build a third dataset using encoders and multiple similarity criterion thresholds. We used the same MPNet threshold and relaxed the USE threshold to $0.50$. Given the differences between the two encoders, we consider a paragraph similar if the difference is at most $0.20$; otherwise, we consider them dissimilar and discard the paragraph. In doing so, we favored the consistency between the two encoding techniques to alleviate the incompatibility of the two encoders.

Some hyperlinks processing did not terminate. Upon examining the content of those pages, we found that they contain a history of software vulnerability with updates and over 20,000 paragraph tags in some cases. Moreover, most of them were not considered by the sentence encoder because they did not meet the threshold.
To address this issue, we consider the first 100 paragraph tags in each hyperlink, as most pages contain related textual information at the top, with subsequent paragraphs reiterating previously mentioned information. 
We consider hyperlinks with valid SSL certificates, limiting our collection to authentic content. Table~\ref{table:datasets} shows the datasets and the number of vulnerabilities using the above method. 

\begin{table}[t]
\centering
\begin{tabular}{l|lr}
  \hline
    \# CVEs & Encoder & Vulnerabilities \\\hline
    \multirow{3}{*}{35,657} & USE & 9,955  \\
    \cline{2-3}
    & MPNet & 8,664  \\
    \cline{2-3}
    & Both & 10,766 \\ \hline
  \end{tabular}
  \caption{Datasets and their high-level characteristics.}\label{table:datasets}\vspace{-8mm}
\end{table}

\subsection{Enhancing Augmented Text Quality}\label{subsec:enhance_ds} Our pipeline gathers text from various sources based on embedding similarity, relying on the sentence encoder for inclusion. However, hyperlinks may lead to redundant paragraphs, resulting in long and repetitive augmented text. To address this, we apply word frequency filtering to enhance diversity. We believe the low evaluation scores of the original dataset were due to the quality of the augmented text, and condensing it into a few key sentences could improve performance.

\subsubsection{Word Frequency} {\em Word count} describes a document as a vector of its word frequency. First, we find unique words or lexicons across all documents, then represent the document as a vector of key-value pairs such that the key is the word. The value is its frequency normalized by the number of occurrences across all documents, indicating the probability of that word. Using these vectors, we compute the cosine similarity of two documents based on their words and associated frequency. 
 
\begin{figure}[t]
    \centering
    \includegraphics[width=0.48\textwidth]{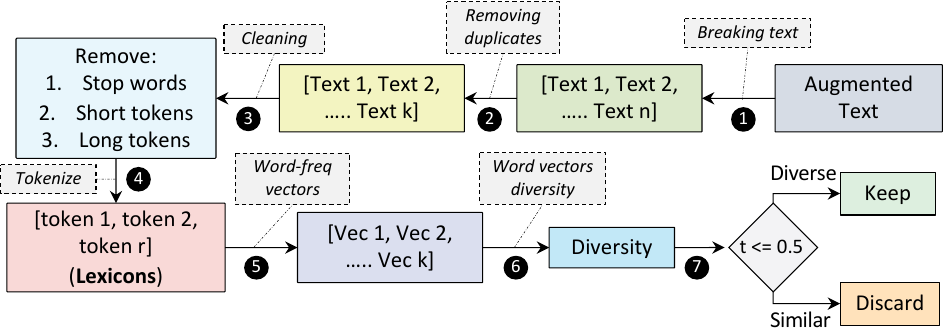}
    \caption{Dataset quality enhancement pipeline, added as an extension to \ours. The pipeline starts with the augmented text of each instance and breaks it into a group of sentences, followed by a duplicate removal pipeline. Text cleaning considers removing ineffective phrases followed by tokenization to construct the lexicons set. Finally, word-frequency vectors are created for each sentence, and the diversity between sentences is measured using a predefined threshold.}
    \label{figure:improving_ds_pipeline}
\end{figure}

\subsubsection{Methodology}  

Before constructing the frequency vector for each sentence, we apply multiple preprocessing steps, as illustrated in Figure~\ref{figure:improving_ds_pipeline}. First, we segment the augmented text into sentences, using a period followed by a space as a separator.  Next, we remove duplicate sentences using the Universal Sentence Encoder (USE). Pairwise similarity is computed between sentences, and those with a similarity score of \(\geq0.98\) are discarded, allowing for minor variations.  We further clean the text by removing stop words, short tokens (fewer than three characters), and long tokens (more than 20 characters), as the latter typically consist of rare character sequences. This step reduces vector space dimensionality while retaining commonly used words relevant to the vulnerability. Finally, we tokenize each sentence and store the tokens in a set representing the lexicon of the augmented text. These tokens are then used to construct frequency vectors for each sentence.

The diversity of two sentences is measured using cosine similarity with a threshold of 0.5, excluding sentences that share up to half of their words. Each new sentence is compared to all others in the set and included only if it meets the diversity requirement. The final sentences form the new augmented text, which is returned to its original form, with any cleaned words restored, and concatenated into one long text using periods as separators. We update the dataset by overwriting the previous augmented text. The pipeline, shown in Figure~\ref{figure:improving_ds_pipeline}, is used with two settings: one without length capping, allowing any text length as long as it meets the conditions, and one with a 250-word cap, chosen because over 83\% and 90\% of the augmented text in the USE and MPNet datasets, respectively, contained 250 words or fewer.

\subsection{Statistical Analysis}

Our datasets are in Table~\ref{table:datasets}. We selected the dataset produced by both encoders for statistical analysis, as it is the largest, to better understand its characteristics. Most augmented descriptions contain \(\leq 1000\) tokens, whereas original summaries typically have \(\leq 200\) tokens. Accordingly, we set training thresholds of 1000 tokens for augmented descriptions and 250 tokens for summaries.  We analyzed word, character, and sentence counts for both augmented text and original summaries, revealing significant differences. The mean and standard deviation were:  (1) Words: (48, 2086) vs. (49, 31). (2) Characters: (2939, 12370) vs. (279, 186) , (3) Sentences: (43, 184) vs. (7, 5.32). These variations highlight the necessity of summarization to normalize the augmented text.

We perform NER to understand which entities were presented across the summary since that is the target in the dataset.
We found the following frequently named entities (with their frequencies): XSS (\textcolor{blue}{799}), N/AC (\textcolor{blue}{523}), IBM X-Force ID (\textcolor{blue}{463}), N/S (\textcolor{blue}{343}), Cisco (\textcolor{blue}{336}), SQL (\textcolor{blue}{334}), Server (\textcolor{blue}{315}), JavaScript (\textcolor{blue}{267}), WordPress (\textcolor{blue}{264}), Jenkins (\textcolor{blue}{240}), IBM (\textcolor{blue}{237}), Firefox (\textcolor{blue}{200}), Java (\textcolor{blue}{187}), VirtualBox (\textcolor{blue}{174}), PHP (\textcolor{blue}{164}), Java SE (\textcolor{blue}{150}), and Android (\textcolor{blue}{148}). The common names include organizations, e.g., Cisco and IBM, technologies, e.g., JavaScript and PHP, or vulnerabilities, e.g., XSS.

We analyzed the most frequent trigrams and found their descriptions meaningful, forming the basis for an acceptable summary. Unlike the augmented text trigrams, which appear uninformative due to repetitiveness, these results further accentuate the need to summarize the augmented text.

\begin{table}[t]
  \caption{Models performance upon fine-tuning reported in terms of recall (R), precision (P), F1-score (F1), text length ($\ell$), beam size ($b$), and batch size (B).}\label{table:comp-results}\vspace{-4mm}
\centering
  \begin{tabular}{L{0.05\textwidth}R{0.05\textwidth}R{0.05\textwidth}R{0.05\textwidth}R{0.05\textwidth}rr}
  \toprule
    Model & R & P & F1 & $\ell$ & $b$ & B\\
  \hline
    \multirow{6}{*}{BART} & 0.51 & 0.50 & 0.49 & 1000 & 2 & 8 \\
    & 0.51 & 0.46 & 0.47 & 1000 & 5 & 8  \\
    & 0.52 & \textbf{0.52} & \textbf{0.51} & 500 & 2 & 8 \\
    & \textbf{0.53} & 0.50 & 0.50 & 500 & 5 & 8 \\
    & 0.50 & 0.51 & 0.49 & 500 & 2 & 4 \\
    & 0.51 & 0.49 & 0.49 & 500 & 5 & 4 \\
    \hline
    \multirow{4}{*}{T5} & 0.46 & 0.50 & 0.47 & 500 & 2 & 8 \\
    &  0.47 & 0.49 & 0.47 & 500 & 5 & 8  \\
    & 0.47 & \textbf{0.52} & \textbf{0.48} & 500 & 2 & 4  \\
    & \textbf{0.47} & 0.50 & 0.47 & 500 & 5 & 4  \\
    \hline
  \end{tabular} \vspace{-7mm}
\end{table}

\subsection{Experimental Settings}
To conduct our experiments, we utilized Pytorch {\tt v1.12} and Pytorch lightning {\tt v1.9} frameworks for training both models. Sklearn {\tt v1.0.2} and Pandas {\tt v1.3} libraries were used for data preprocessing, while spacy {\tt v3.4.1} was used for NER, and NLTK {\tt v3.7} was used for text preprocessing. We deployed the models using the transformers library {\tt v.4.10} from Hugging Face for BART and T5. The experiments were run on a cloud GPU with a Tesla P100 and 16 GB of memory.

We split the dataset, reserving 10\% for testing, then further split the training set with 10\% for validation. The models are trained for 4 epochs with a batch size of 8 and a learning rate of 0.0001, based on parameter testing (results omitted for space). Beam search is used for decoding with a beam size of 2. We also fix several parameters: a length penalty of 8 to encourage longer summaries and a repetition penalty of 2 to avoid repeating generated words. These values were selected based on experiments yielding the best results.

\subsection{Computational Metrics and Results}
ROUGE measures the matching n-gram between the prediction and the target.
For our evaluation, we use ROUGE-1, which measures the overlapping unigrams and gives an approximation of the overlap based on individual words.
ROGUE consists of three sub-metrics: recall, precision, and F1-score. The recall measures the number of matching n-grams between the generated and the target summary, normalized by the number of words in the target summary.
In contrast, precision normalizes the same quantity as in the recall but by the number of words in the generated summary.
Finally, the harmonic mean or F1-score is expressed as $ F1{-}Score = 2\times ({precision \times{recall}})/({precision + {recall}})$.

Table~\ref{table:comp-results} shows the ROUGE scores after fine-tuning BART and T5. We began with BART due to its ease and speed of training, requiring fewer resources than T5. We observed that all metrics improved when the augmented text was limited to 500 tokens, and most metrics performed better with smaller beams. This may be because larger beams increase the risk of selecting incorrect sequences. Additionally, more beams lead to longer summary generation times. Based on these initial results with BART and the resource demands of T5, we decided to train T5 on text limited to 500 tokens. We found that a batch size of 4 outperformed a batch size of 8 across all metrics for T5, and increasing the number of beams did not improve its performance.

     \begin{figure}[t]
     \pgfplotsset{width=4cm,height=3cm,compat=1.8}  
       \centering
    \begin{tikzpicture}  
          
        \begin{axis}  
        [  
            ybar, 
            enlargelimits=0.3,  
            ylabel={Metric Avg Value},  
            symbolic x coords={Recall, Precision, F1-score}, 
            xtick=data,  
            x=2cm,
            minor tick num = 3,
            minor grid style = {lightgray!25},
            font=\footnotesize,
            ymin=0.25,
            legend style={font=\footnotesize,at={(-0.1,1.35)},anchor=north west,
                    legend columns=4},
            xticklabel style={rotate=0},
            grid=both,grid style={line width=.1pt, draw=gray!10},major grid style={font=\tiny,line width=.1pt,draw=gray!10}
            ]  
        \addplot[color=black,pattern=grid] coordinates {(Recall,0.61) (Precision,0.6) (F1-score,0.59)};  
        \addlegendentry{BART-USE};
        \addplot[color=black,fill=black,pattern=north east lines] coordinates {(Recall,0.55) (Precision,0.57) (F1-score,0.55)};
        \addlegendentry{BART-MPNet};
        \addplot[color=black,pattern=crosshatch] coordinates {(Recall,0.58) (Precision,0.62) (F1-score,0.59)};  
        \addlegendentry{T5-USE};
        \addplot[color=black, pattern=north west lines] coordinates {(Recall,0.53) (Precision,0.59) (F1-score,0.54)};
        \addlegendentry{T5-MPNet};
        \end{axis}  
        \end{tikzpicture}  \vspace{-3mm}
         \caption{Results after fine-tuning the models using a different single encoder. The number of beams used across all settings is 2, and the batch size for BART is 8, while for T5 is 4. The keys highlight pairs of model-encoder for BART and T5 using both USE and MPNet.}
         \label{table:results-encoders}\vspace{-3mm}
    \end{figure}
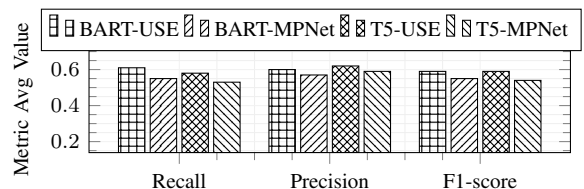

Figure~\ref{table:results-encoders} shows the computational metrics for the two other datasets, which outperformed the mixed dataset, highlighting that using two encoders with different architectures may harm the performance.
While the {\em USE} dataset is larger, we believe the results are better due to {\em USE}'s accuracy in semantics encoding, allowing it to produce a reliable representation of long text.
We reiterate that we used the DAN architecture for USE, which is less accurate than the transformer architecture.

{\noindent\bf\em Summary-Target Comparison:} We compared the target summary with the model-generated summary using the same sentence encoders.
We encoded both summaries (original and new) using both encoders and measured the similarity between the target and the prediction. Most predictions are very close to the target, with the distribution mean around a similarity of $0.75$--omitted for the lack of space. 
This indicates that the models are learning and generating summaries similar to the target in the embedding space of each sentence encoder. 

\subsection{Human Metrics Results}
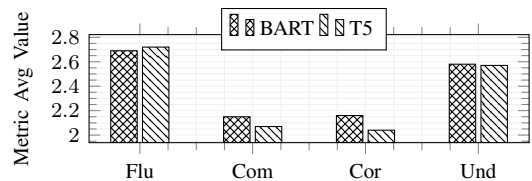
\begin{figure}[t]
\pgfplotsset{width=3.5cm,compat=1.7}  
\centering
    \begin{tikzpicture}  
          
        \begin{axis}  
        [  
            ybar, 
            enlargelimits=0.15,  
            ylabel={Metric Avg Value},  
            symbolic x coords={Flu, Com, Cor, Und}, 
            xtick=data,  
            x=1.5cm,
            minor tick num = 3,
            minor grid style = {lightgray!25},
            font=\footnotesize,
            legend style={at={(0.3,1.05)},anchor=west}, legend columns=2,
            xticklabel style={rotate=0},
            grid=both,grid style={line width=.1pt, draw=gray!10},major grid style={line width=.1pt,draw=gray!10}
            ]  
        \addplot[color=black,pattern=crosshatch] coordinates {(Flu,2.69) (Com,2.15) (Cor,2.16) (Und,2.58)};  
        \addlegendentry{BART};
        \addplot[color=black, pattern=north west lines]  coordinates {(Flu,2.72) (Com,2.07) (Cor,2.04) (Und,2.57)};  
         \addlegendentry{T5};
        \end{axis}  
        \end{tikzpicture} \vspace{-3mm} 
         \caption{Evaluation: \underline{Flu}ency, \underline{Com}pleteness, \underline{Cor}rectness, and \underline{Und}erstanding.}
         \label{table:HE}\vspace{-3mm}
\end{figure}

We used four human metrics: fluency, correctness, completeness, and understanding.
These metrics are graded on a discrete scale between 1 and 3--1 is the lowest, and 3 is the highest. The human metrics are defined as follows: 
\begin{itemize}
    \item {\em Fluency.} This metric measures the generated summary's grammatical structure and semantic coherence. 
    \item {\em Correctness.} This metric measures the model prediction accuracy in capturing the correct vulnerability details. 
\item {\em Completeness.} This metric measures how complete the generated summary is for details in the target summary.
\item {\em Understanding.} This metric measures how easy it is to understand the generated summary.
\end{itemize}
The evaluation is performed over 100 randomly selected samples where we report the average results in Figure~\ref{table:HE}.  
    
\BfPara{Quantitative Results} 
Our evaluation showed that both models produced fluent and easy-to-understand summaries with few exceptions. Understanding scored highly, as it is influenced by fluency, meaning the summaries were mostly cohesive and clear. However, short summaries provided limited context. In contrast, completeness and correctness suffered in both models, as \ours was not designed to excel in these areas. Additionally, the varying lengths of the augmented text may have impacted the performance metrics. Nevertheless, these two metrics performed well when the augmented text had a specific length, capturing most details. Both models were comparable in terms of human metrics when their generated summaries were compared against the target.

\BfPara{Qualitative Results} Both models exhibited unpredictable behaviors, such as repeating sentences or software names multiple times and introducing unrelated software into predictions. They also tended to be extractive when the augmented text ranged between 70-130 words. The length of the augmented description significantly impacted predictions. With very short text ($\sim$20 words), models generated summaries learned during training, sometimes including common phrases like ``gain access'' or ``code execution'' even if absent from the input. Conversely, when the text was too long ($\geq$260 words), predictions became repetitive but still covered key aspects of the summary. Other observed issues included the use of opposite adjectives and missing software names, even when capitalized in the augmented text. These defects appeared in both models to varying degrees. A potential solution is ensuring greater diversity in augmented sentences and removing repetitions.

\subsection{Analysis of the Summary Contents}
We performed additional post-analysis on the generated summary and compared it to the original summary to understand the effect of our approach.
We selected BART-generated summaries trained on the USE dataset because the model achieved the best computational metrics results.
We did not rely on human metrics here because (1) they are subjective, and (2) the evaluation considered only 100 samples.

\begin{figure}[t]
\pgfplotsset{width=4cm,height=3cm,compat=1.9}  
       \centering
    \begin{tikzpicture}  
          
        \begin{axis}  
        [  
            ybar, 
            enlargelimits=0.07,  
            ylabel={\# of Samples}, 
            xlabel={\# of Tokens},
            bar width=3mm,
            symbolic x coords={0-25, 26-50, 51-75, 76-100, 101-125, 126-150, 151-175,176-200, >200}, 
            xtick=data,  
            x=0.8cm,
            minor tick num = 3,
            minor grid style = {lightgray!25},
            font=\footnotesize,
            legend style={at={(0.3,1.05)},anchor=west, legend columns=2},
            xticklabel style={rotate=20,anchor=east, align=right},
            grid=both,grid style={line width=.1pt, draw=gray!10},major grid style={line width=.1pt,draw=black!10}
            ]
        \addplot[color=black,pattern=crosshatch]  coordinates {(0-25,13) (26-50,295) (51-75,309) (76-100,148) (101-125,111) (126-150,31) (151-175,26) (176-200,43) (>200,20)};  
         \addlegendentry{Generated};
        \addplot[color=black, pattern=north west lines] coordinates {(0-25,19) (26-50,266) (51-75,248) (76-100,163) (101-125,120) (126-150,55) (151-175,51) (176-200,33) (>200,41)};  
        \addlegendentry{Original};
        \end{axis}  
        \end{tikzpicture}  \vspace{-3mm}
    \caption{Tokens distribution of the original and generated summaries.}
    \label{figure:Word_dist}\vspace{-3mm}
    \end{figure}
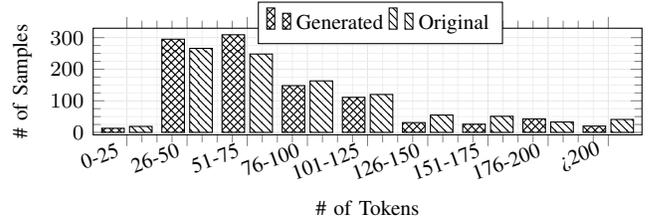

First, we expected restricting the generated summary length to $250$ tokens during fine-tuning would yield a uniform summary, achieving the maximum limit for most instances by considering the lengthy augmented text.  
However, most generated summaries were significantly below $250$ tokens, as shown in Figure~\ref{figure:Word_dist}. 
We see in Figure~\ref{figure:Word_dist} that the number of samples with token counts between 26 and 75 has increased. Still, as the number of tokens starts to increase, the number of instances decreases in the generated summary, even more than the original summary, except for the token counts between 176 and 200.
Moreover, the statistics for the generated summary (mean and standard deviation) were also lower than the original summary, which goes against our expectations.

     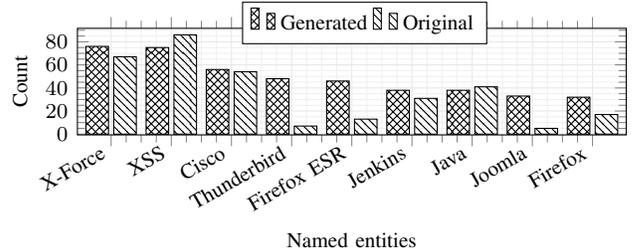
\begin{figure}[t]
\pgfplotsset{width=4cm,height=3cm,compat=1.5}  
       \centering
    \begin{tikzpicture}  
          
        \begin{axis}  
        [  
            ybar, 
            enlargelimits=0.07,  
            ylabel={Count}, 
            xlabel={Named entities},
            symbolic x coords={X-Force, XSS, Cisco, Thunderbird, Firefox ESR, Jenkins, Java,Joomla, Firefox}, 
            xtick=data,  
            minor tick num = 3,
            minor grid style = {lightgray!25},
            font=\footnotesize,
            x=0.8cm,
            bar width=3mm,
            legend style={at={(0.3,1.05)},anchor=west, legend columns=2},
            xticklabel style={rotate=30,anchor=east, align=right},
            grid=both,grid style={line width=.1pt, draw=gray!10},major grid style={line width=.1pt,draw=black!10}
            ]
        \addplot[color=black,pattern=crosshatch]  coordinates {(X-Force,76) (XSS,75) (Cisco,56) (Thunderbird,48) (Firefox ESR,46) (Jenkins,38) (Java,38) (Joomla,33) (Firefox,32)};  
         \addlegendentry{Generated};
        \addplot[color=black, pattern=north west lines] coordinates {(X-Force,67) (XSS,86) (Cisco,54) (Thunderbird,7) (Firefox ESR,13) (Jenkins,31) (Java,41) (Joomla,5) (Firefox,17)};  
        \addlegendentry{Original};
        \end{axis}  
        \end{tikzpicture}  \vspace{-3mm}
    \caption{The top names in the original and generated summary.}
    \label{figure:NER}\vspace{-3mm}
    \end{figure}
    
We analyzed the NER shown in Figure~\ref{figure:NER} and found that many entities appear in both summaries, although their counts did not align. 
XSS appeared to be the most repeated name in the original summary but came second in the generated summary.
We also noticed that some names, such as Thunderbird and Firefox ESR, appeared many times in the generated summary but were not present in the original summary, which could indicate overfitting.
This result corroborates what was reported in our human evaluation, where predictions included software not mentioned in the original summary.

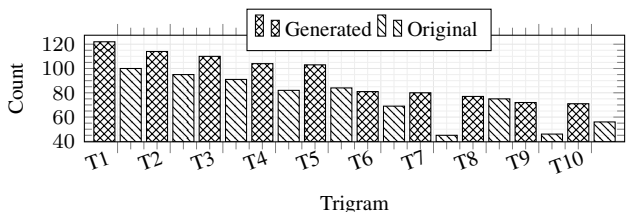
\begin{figure}[t]
\pgfplotsset{width=4cm,height=3cm,compat=1.5}  
       \centering
    \begin{tikzpicture}  
          
        \begin{axis}  
        [  
            ybar, 
            enlargelimits=0.07,  
            ylabel={Count}, 
            xlabel={Trigram},
            symbolic x coords={T1, T2, T3, T4, T5, T6, T7,T8, T9, T10}, 
            xtick=data,  
            x=0.7cm,
            bar width=2.8mm,
            minor tick num = 3,
            minor grid style = {lightgray!25},
            font=\footnotesize,
            legend style={at={(0.3,1.05)},anchor=west, legend columns=2},
            xticklabel style={rotate=20,anchor=east, align=right},
            grid=both,grid style={line width=.1pt, draw=gray!10},major grid style={line width=.1pt,draw=black!10}
            ]
        \addplot[color=black,pattern=crosshatch]  coordinates {(T1,122) (T2,114) (T3,110) (T4,104) (T5,103) (T6,81) (T7,80) (T8,77) (T9,72) (T10,71)};  
         \addlegendentry{Generated};
        \addplot[color=black, pattern=north west lines] coordinates {(T1,100) (T2,95) (T3,91) (T4,82) (T5,84) (T6,69) (T7,45) (T8,75) (T9,46) (T10,56)};  
        \addlegendentry{Original};
        \end{axis}  
        \end{tikzpicture}  \vspace{-5mm}
    \caption{Trigrams count. The generated summary has the same trigram repeated more than the original.}
    \label{figure:trigram}\vspace{-3mm}
    \end{figure}

We analyzed the set of the most repeated trigrams produced by the model, T1 through T10, defined as: 
{\footnotesize
\begin{tabular}{llll}
T1 & ``could allow attacker'' & 
T2 & ``attacker could exploit''\\
T3& ``could exploit vulnerability''& 
T4& ``successful exploit could''\\
T5& ``exploit could allow''&
T6& ``execute arbitrary code''\\
T7& ``attacker executes arbitrary''&
T8& ``IBM X-Force id''\\
T9& ``exploit vulnerability sending''&
T10& ``supported version affected''\\
\end{tabular}
}
The frequency is shown in  Figure~\ref{figure:trigram}.
Although the model produced meaningful and informative trigrams like the original summary, the generated summary exceeded the original, which is evident for every trigram.
These results indicate that the model is overfitting as it memorizes what it learned during training and uses it for inferences, even when unrelated. 
One explanation is the dataset quality, as the augmented text includes repeated content, thus this behavior. 

Our results indicate that further improvements are needed for generating high-quality summaries. Notably, larger models do not always perform better, especially on curated datasets tailored to specific domains. Additionally, we prioritized hyperparameter fine-tuning while overlooking potential dataset refinements.  Overfitting was evident when training loss decreased while validation loss increased. However, during fine-tuning, validation loss consistently decreased while training loss fluctuated, likely due to varying lengths of augmented text. In our case, overfitting was identified through human evaluation and post-statistical analysis.

\subsection{Analysis of the Label-Guided Dataset}
We report our results using the label-guided dataset in section~\ref{sec:dataset}. To make our evaluation concrete, we fine-tuned both models using the same $100$ samples but with the original summary as the target to compare both approaches.
Considering computational metrics, we noticed that the label-guided approach outperformed the description-guided approach, as shown in Table~\ref{table:comp-results-label_guided}. This is expected since our ground-truth summaries are created from the augmented text, confirming that using the description as a summary is insufficient.
We hypothesize this is due to the difference in length and content between the augmented text and the description. In contrast, our summaries provide better and longer context to the model, generating coherent and plausible summaries. 

\begin{figure}[t]
\pgfplotsset{width=4cm,height=3cm,compat=1.7}  
       \centering
    \begin{tikzpicture}  
          
        \begin{axis}  
        [  
            ybar, 
            enlargelimits=0.15,  
            ylabel={Metric Avg Value},  
            symbolic x coords={\underline{Flu}, \underline{Com}, \underline{Cor}, \underline{Und}}, 
            xtick=data,  
            x=1.5cm,
            minor tick num = 3,
            minor grid style = {lightgray!25},
            font=\footnotesize,
            legend style={at={(0.3,1.05)},anchor=west, legend columns=2},
            xticklabel style={rotate=0},
            grid=both,grid style={line width=.1pt, draw=gray!10},major grid style={line width=.1pt,draw=black!10}
            ]  
        \addplot[color=black,pattern=crosshatch] coordinates {(\underline{Flu},2.6) (\underline{Com},2.55) (\underline{Cor},2.25) (\underline{Und},2.45)};  
        \addlegendentry{BART};
        \addplot[color=black, pattern=north west lines]  coordinates {(\underline{Flu},2.65) (\underline{Com},2.6) (\underline{Cor},1.8) (\underline{Und},1.95)};  
         \addlegendentry{T5};
        \end{axis}  
        \end{tikzpicture}  \vspace{-3mm}
         \caption{The human evaluation metric results using the label-guided approach: \underline{Flu}ency, \underline{Com}pleteness, \underline{Cor}rectness, and \underline{Und}erstanding.}
         \label{table:HE-label_guided}\vspace{-5mm}
    \end{figure}
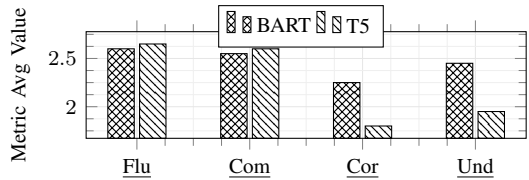

For the human metrics in Figure~\ref{table:HE-label_guided}, both models' fluency and completeness scores are relatively higher than the correctness and understanding scores, especially for T5. 
Moreover, we found that both models generated sound and fluent summaries, including most details from the target summaries. 
In contrast, correctness is the most affected metric as the details included were incorrect (e.g., wrong software version or bug description) from the created summary. The understanding was problematic because the generated summary needed more context, or the generated sentences were sparse. 

Both models generated summaries by extracting salient spans from the augmented text but did not align with hand-crafted summaries. Software names in labels were often missing, and T5 sometimes copied the initial text spans, producing fluent yet unclear summaries. Despite limitations, the results suggest that using descriptions as summaries is insufficient.

\begin{table}[t]
\centering
  \begin{tabular}{llccccc}
        \rotatebox[origin=c]{0}{{Model}} & 
        \rotatebox[origin=c]{0}{{Dataset}} & \rotatebox[origin=c]{60}{{Recall}} & \rotatebox[origin=c]{60}{{Precision}} & \rotatebox[origin=c]{60}{{F1-score}} & \rotatebox[origin=c]{60}{{Batch}}\\
  \Xhline{1\arrayrulewidth}
    \multirow{4}{*}{BART} & Ground-truth & 0.59 & 0.55 & 0.54 & 4 \\
    & Description & 0.36 & 0.35 & 0.33 & 4  \\
    \cline{2-6}
    & Ground-truth & 0.61 & 0.55 & 0.55 & 8  \\
    & Description & 0.32 & 0.33 & 0.29  & 8  \\
    \Xhline{1\arrayrulewidth}
    \multirow{2}{*}{T5} & Ground-truth & 0.49 & 0.58 & 0.51 & 4 \\
    & Description & 0.30 & 0.32 & 0.30 & 4  \\
    \Xhline{1\arrayrulewidth}
  \end{tabular}
    \caption{Results after fine-tuning the models using the label-guided and the description-guided methods.}
  \label{table:comp-results-label_guided}\vspace{-5mm}
\end{table}

\subsection{Analysis of Quality Enhancement Pipeline}
Considering our initial results on the three datasets, we applied our enhancement pipeline to the USE dataset because it achieved the best results. 
The dataset is fed into the pipeline, as shown in Figure~\ref{figure:improving_ds_pipeline}, and processed accordingly.
Capping, by limiting the length to 250 words in the augmented text, is applied after ensuring the diversity of a sentence. In particular, we check each sentence length to ensure the total length is $\leq 250$ words. 
Our goal of capping is to find out whether limiting the length will help the model.  
We obtain two new datasets after passing the USE dataset through the pipeline.
Due to our pipeline in Figure \ref{figure:improving_ds_pipeline}, which ensures preserving a set of diverse sentences, both datasets' sizes have shrunk by almost 50\% of their original size.  We used the same setting reported when using the USE dataset as shown in Figure~\ref{table:results-encoders}. 

\begin{figure}[t]
     \pgfplotsset{width=4cm,compat=1.8} 
       \centering
    \begin{tikzpicture}  
          
        \begin{axis}  
        [  
            ybar, 
            enlargelimits=0.3,  
            ylabel={Metric Avg Value},  
            symbolic x coords={Recall, Precision, F1-score}, 
            xtick=data,  
            x=2cm,
            font=\footnotesize,
            ymin=0.25,
            minor tick num = 3,
            minor grid style = {lightgray!25},
            legend style={at={(0.10,1.38)},anchor=north west,
                    legend columns=2},
            xticklabel style={rotate=0},
            grid=both,grid style={line width=.1pt, draw=gray!10},major grid style={line width=.1pt,draw=gray!10}
            ]  
        \addplot[color=black,pattern=grid] coordinates {(Recall,0.62) (Precision,0.59) (F1-score,0.59)};  
        \addlegendentry{BART-Capped};
        \addplot[color=black,fill=black,pattern=north east lines] coordinates {(Recall,0.61) (Precision,0.61) (F1-score,0.59)};
        \addlegendentry{BART-Uncapped};
        \addplot[color=black,pattern=crosshatch] coordinates {(Recall,0.57) (Precision,0.63) (F1-score,0.59)};  
        \addlegendentry{T5-Capped};
        \addplot[color=black, pattern=north west lines] coordinates {(Recall,0.57) (Precision,0.63) (F1-score,0.59)};
        \addlegendentry{T5-Uncapped};
        \end{axis}  
        \end{tikzpicture} \vspace{-3mm} 
         \caption{Results after fine-tuning the models after passing the USE dataset through the quality enhancement pipeline. The number of beams used across all settings is 2, and the batch size for BART is 8, while for T5 is 4.}
         \label{table:results-enhanced}\vspace{-3mm}
    \end{figure}
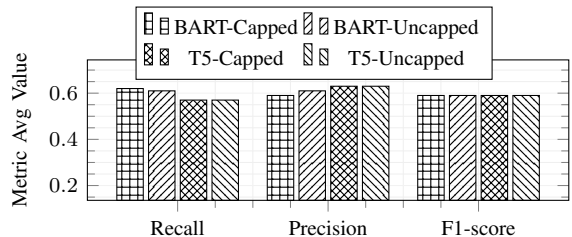

The results for both models are presented in Figure~\ref{table:results-enhanced}. We observed that T5 produced identical results across both datasets, consistent with Figure~\ref{table:results-encoders}. In contrast, BART showed a slight improvement in recall when applied to the capped dataset but no significant change in precision.   Despite the dataset shrinking by nearly 50\%, computational metrics remained unchanged, suggesting that transformer models can recognize similar texts and may not focus heavily on fine-tuning specific tasks. This aligns with the understanding that pre-trained language models are trained on vast corpora, which often include duplicates or highly similar entries. Additionally, the diversity pipeline alters sentence flow in the augmented text, potentially reducing coherence.   However, we observed a decrease in validation loss for both models. Additionally, T5 showed a reduction in training loss, indicating improved learning efficiency with the refined dataset.

The training and validation loss dropped, proving the quality enhancement pipeline improved the dataset, and its effect can be seen in training. For BART, the validation loss decreased from $0.46$ to $0.40$. For T5, the training and validation loss decreased from $2.35$ and $1.46$ to $0.99$ and $1.20$, respectively. 

Relying solely on the ROUGE score is insufficient, as it only measures word overlap. The original dataset, which contains more text, forces the model to generate summaries with a wider vocabulary, increasing the overlap. However, maintaining the same score while reducing the dataset size is promising, as data reduction typically degrades performance.   To further evaluate summary quality, we analyzed 100 randomly selected instances. The results showed noticeable improvement. Repetitive behavior, previously observed in seven samples, was rare in the new summaries. Additionally, the generated summaries were better aligned with target summaries in content and maintained a consistent length.

The model learned to map some concepts, e.g., the target summary included XSS, but the generated summary uses cross-site scripting (XSS).
Moreover, the generated summary is more relevant to the target software name but is still inaccurate.
However, the generated summary still suffers from adding unrelated software, but not as much as it did with the original dataset.
Overall, our enhancement pipeline has improved the dataset's quality by reducing its size, the loss for training and validation, and the quality of the generated summaries.
However, we believe additional steps are needed to address some deficiencies in the created dataset. 
For example, ensuring the content is closely related and constitutes useful detail to the vulnerability may require a more sophisticated approach instead of simple similarity criteria, especially since we are scraping blindly from third-party resources.  

\subsection{Case Study}
In this section, we use \textit{CVE-2019-5005} as a case study to demonstrate Zad's effectiveness by showing the original description against the enhanced and enriched description. 

\textbf{Original Description:} An issue was discovered in Foxit Reader and PhantomPDF before 9.4 on Windows. They allowed Denial of Service (application crash) via image data because two bytes are written to the end of the allocated memory without judging whether this will cause corruption. 
\textbf{Zad’s summary:} This vulnerability allows remote attackers to execute arbitrary code on vulnerable Foxit Reader and PhantomPDF installations. User interaction is required to exploit this vulnerability in that the target must visit a malicious page or open a malicious file. A specific flaw exists within the processing of PDF files. The issue results from the lack of proper validation of user-supplied data, which can result in a read past the end of an allocated buffer. An attacker can leverage this in conjunction with other vulnerabilities to execute code in the context of the current process.

The key enhancements provided by Zad included crucial details such as user interaction requirements for exploitation, which necessitates the target to visit a malicious page or open a malicious file. Moreover, It dives deeper into the technical mechanics of the vulnerability, pinpointing the issue to inadequate validation of user-supplied data during PDF file processing, leading to a buffer read overflow.  Moreover, Zad sheds light on the critical risk of remote attackers executing arbitrary code within the current process context. Although Zad may not capture every technical detail from the original description, the depth and clarity it adds make it a valuable tool that aids cybersecurity professionals.

\section{User study}
Our user study assesses  \ours{'s} effectiveness. The study presented participants with the original description of a vulnerability and the enriched description generated by \ours. The participants were asked to rate the enriched description on enrichment, accuracy, and understanding. Each criterion was rated on a scale of 1 (worst) to 3 (best).

{\em Enrichment} refers to how the enriched description improved upon the original with context and related information. A rating of 3 was given if the generated description included more details than the original, two if the two descriptions were relatively the same, and one if the enriched description had less content than the original.
{\em Accuracy} measures the ability of the enriched description to capture the software and bug information in the original description. A rating of 3 was given if the generated description included all the details, 2 if it missed at most two pieces of information, and 1 if it missed most details.
Finally, understanding refers to the user's ability to comprehend the text's meaning. A rating of 3 was given if the meaning could be easily understood, 2 if it was more difficult but still possible, and 1 if it could not be understood.

We selected BART as the most effective model and used the USE dataset. To evaluate its performance, we randomly sampled 100 enriched descriptions generated by the model alongside their original descriptions. Due to constraints in participant recruitment and resource availability, we limited the experiment to five evaluators, all experts in threat intelligence and security operations. The results were consistent among evaluators, especially for enrichment and understanding, with enrichment scores ranging from 2.53 to 2.60 (median: 2.55), confirming that the summaries provided more context than the originals. The extended descriptions also improved understanding, with scores between 2.70 and 2.87 (median: 2.78). The strong correlation between these metrics highlights \ours's effectiveness in enhancing vulnerability content.  However, accuracy varied among evaluators, averaging between 2.36 and 2.74 (median: 2.53), revealing \ours's limitations. Capturing software and bug details requires support from an additional model. Since \ours primarily focuses on improving description sufficiency, further refinement is needed to enhance accuracy in identifying vulnerability-specific details.

 \begin{figure}
        \centering\pgfplotsset{width=8cm,height=3cm,compat=1.18}
        \begin{tikzpicture}
        \begin{axis}[
            boxplot/draw direction=y,
            enlarge y limits,
            font=\footnotesize,
            ymajorgrids,
            minor tick num = 3,
            minor grid style = {lightgray!25},
            xtick={1,2,3},
            xticklabels={Enrichment, Accuracy, Understanding},
        ]
        \addplot+ [
        fill,fill opacity=0.5,color=black,
        boxplot prepared={
        lower whisker=2.53, lower quartile=2.535,
        median=2.55,
        upper quartile=2.58, upper whisker=2.60,
        box extend=0.3},
        ] coordinates {};
        
        \addplot+ [
        fill,fill opacity=0.5,color=black,
        boxplot prepared={
        lower whisker=2.36, lower quartile=2.4,
        median=2.51,
        upper quartile=2.675, upper whisker=2.74,
        box extend=0.3
        },
        ] coordinates {};        
        
        \addplot+ [
        fill,fill opacity=0.5,color=black,
        boxplot prepared={
        lower whisker=2.7, lower quartile=2.73,
        median=2.83,
        upper quartile=2.85, upper whisker=2.87,
        box extend=0.3
        },
        ] coordinates {};
        \end{axis}
        \end{tikzpicture}
        \caption{User study evaluation over 100 samples generated by BART using the USE dataset with five evaluators.}\vspace{-5mm}
    \end{figure}
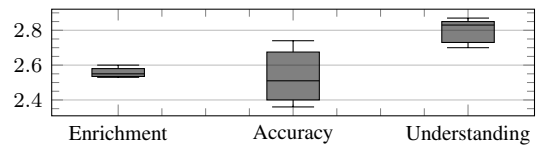

The overall positive response from the enriched descriptions generated by Zad underscores its effectiveness in demystifying complex cybersecurity challenges. By extracting and integrating information from dense technical reports into easily understandable vulnerability descriptions, Zad serves as a pivotal educational tool and demonstrates its value beyond cybersecurity. Zad’s methodology establishes the essence of information retrieval to enhance its quality, proving its utility across a spectrum of fields. 

\section{Conclusion}\label{sec:conclusion}
We introduced \ours, a multi-stage pipeline that enhances brief vulnerability descriptions in public databases by leveraging third-party reports. By augmenting and summarizing these reports, \ours generates comprehensive, context-rich descriptions. It integrates security community insights to create well-rounded summaries while employing an efficient approach to improve dataset quality.  Although \ours effectively enriches descriptions with semantically relevant content, improving content relatedness remains a challenge. Future work could involve training a classification model to refine dataset filtering or employing contrastive learning to strengthen associations between related texts. Such augmentation-focused models could further enhance \ours, producing a more concise and informative dataset for better summary generation.


\begin{thebibliography}{10}

\bibitem{bugtraq}
{---}.
\newblock \href{https://bugtraq.securityfocus.com/}{BugTraq by Accenture
  Security}.
\newblock Online, 2021.

\bibitem{MPnet_se}
{---}.
\newblock
  \href{https://huggingface.co/sentence-transformers/all-mpnet-base-v2}{all-mpnet-base-v2}.
\newblock Online, June 2021.

\bibitem{web_ref4}
{---}.
\newblock \href{https://cve.mitre.org/}{MITRE's Common Vulnerabilities and
  Exposures (CVE)}.
\newblock Online, 2022.

\bibitem{web_ref5}
{---}.
\newblock \href{https://nvd.nist.gov/}{National Vulnerability Database (NVD)}.
\newblock Online, 2022.

\bibitem{alabduljabbar2021tldr}
A.~Alabduljabbar, A.~Abusnaina, {\"{U}}.~Meteriz{-}Yildiran, and D.~Mohaisen.
\newblock \href{https://doi.org/10.1145/3463676.3485608}{{TLDR:} Deep
  Learning-Based Automated Privacy Policy Annotation with Key Policy
  Highlights}.
\newblock In {\em ACM WPES}, pages 103--118, 2021.

\bibitem{abs-2210-01260}
H.~Althebeiti and D.~Mohaisen.
\newblock Enriching vulnerability reports through automated and augmented
  description summarization.
\newblock {\em CoRR}, abs/2210.01260, 2022.

\bibitem{AnwarACLM22}
A.~Anwar, A.~Abusnaina, S.~Chen, F.~Li, and D.~Mohaisen.
\newblock Cleaning the {NVD:} comprehensive quality assessment, improvements,
  and analyses.
\newblock {\em {IEEE} Trans. Dependable Secur. Comput.}, 19(6):4255--4269,
  2022.

\bibitem{anwar2020measuring}
A.~Anwar, A.~Khormali, J.~Choi, H.~Alasmary, S.~Salem, D.~Nyang, and
  D.~Mohaisen.
\newblock \href{https://doi.org/10.4108/eai.13-7-2018.164551}{Measuring the
  Cost of Software Vulnerabilities}.
\newblock {\em Transactions on Security and Safety}, 7(23):e1, 2020.

\bibitem{AnwarKNM18}
A.~Anwar, A.~Khormali, D.~Nyang, and A.~Mohaisen.
\newblock \href{https://doi.org/10.1007/978-3-030-01701-9\_21}{Understanding
  the Hidden Cost of Software Vulnerabilities: Measurements and Predictions}.
\newblock In {\em EAI {SecureComm}}, volume 254, pages 377--395, 2018.

\bibitem{USE}
D.~Cer, Y.~Yang, S.~Kong, N.~Hua, N.~Limtiaco, R.~S. John, N.~Constant,
  M.~Guajardo{-}Cespedes, S.~Yuan, C.~Tar, B.~Strope, and R.~Kurzweil.
\newblock \href{https://doi.org/10.18653/v1/d18-2029}{Universal Sentence
  Encoder for English}.
\newblock In {\em {EMNLP}}, pages 169--174, 2018.

\bibitem{ChenKMAA20}
J.~Chen, P.~K. Kudjo, S.~Mensah, S.~B. Aformaley, and G.~Akorfu.
\newblock An automatic software vulnerability classification framework using
  term frequency-inverse gravity moment and feature selection.
\newblock {\em J. Syst. Softw.}, 167:110616, 2020.

\bibitem{BERT}
J.~Devlin, M.~Chang, K.~Lee, and K.~Toutanova.
\newblock \href{https://doi.org/10.18653/v1/n19-1423}{{BERT:} Pre-training of
  Deep Bidirectional Transformers for Language Understanding}.
\newblock In {\em {NAACL-HLT}}, pages 4171--4186, 2019.

\bibitem{ref_proc1}
Y.~Dong, W.~Guo, Y.~Chen, X.~Xing, Y.~Zhang, and G.~Wang.
\newblock
  \href{https://www.usenix.org/conference/usenixsecurity19/presentation/dong}{Towards
  the Detection of Inconsistencies in Public Security Vulnerability Reports}.
\newblock In {\em {USENIX} Security Symposium}, pages 869--885, 2019.

\bibitem{gage1994new}
P.~Gage.
\newblock A new algorithm for data compression.
\newblock {\em C Users Journal}, 12(2):23--38, 1994.

\bibitem{FFN_mem}
M.~Geva, R.~Schuster, J.~Berant, and O.~Levy.
\newblock \href{https://doi.org/10.18653/v1/2021.emnlp-main.446}{Transformer
  Feed-Forward Layers Are Key-Value Memories}.
\newblock In {\em {EMNLP}}, pages 5484--5495, 2021.

\bibitem{guo2020predicting}
H.~Guo, Z.~Xing, and X.~Li.
\newblock \href{https://doi.org/10.1145/3366424.3382707}{Predicting Missing
  Information of Vulnerability Reports}.
\newblock In {\em WWW}, pages 81--82, 2020.

\bibitem{ulmfit}
J.~Howard and S.~Ruder.
\newblock \href{https://aclanthology.org/P18-1031/}{Universal Language Model
  Fine-tuning for Text Classification}.
\newblock In {\em {ACL}}, pages 328--339, 2018.

\bibitem{ibm}
IBM.
\newblock Ibm security x-force.
\newblock {\url{https://www.ibm.com/security/xforce}}, 2022.

\bibitem{DAN}
M.~Iyyer, V.~Manjunatha, J.~L. Boyd{-}Graber, and H.~D. III.
\newblock \href{https://doi.org/10.3115/v1/p15-1162}{Deep Unordered Composition
  Rivals Syntactic Methods for Text Classification}.
\newblock In {\em {ACL}}, pages 1681--1691, 2015.

\bibitem{ref_article1}
K.~Kanakogi, H.~Washizaki, Y.~Fukazawa, S.~Ogata, T.~Okubo, T.~Kato, H.~Kanuka,
  A.~Hazeyama, and N.~Yoshioka.
\newblock \href{https://doi.org/10.3390/info12080298}{Tracing {CVE}
  Vulnerability Information to {CAPEC} Attack Patterns Using Natural Language
  Processing Techniques}.
\newblock {\em Information}, 12(8):298, 2021.

\bibitem{ref_proc3}
K.~Kanakogi, H.~Washizaki, Y.~Fukazawa, S.~Ogata, T.~Okubo, T.~Kato, H.~Kanuka,
  A.~Hazeyama, and N.~Yoshioka.
\newblock \href{https://hdl.handle.net/10125/71462}{Tracing {CAPEC} Attack
  Patterns from {CVE} Vulnerability Information using Natural Language
  Processing Technique}.
\newblock In {\em {HICSS}}, pages 1--9, 2021.

\bibitem{krizhevsky2017imagenet}
A.~Krizhevsky, I.~Sutskever, and G.~E. Hinton.
\newblock
  \href{https://proceedings.neurips.cc/paper/2012/hash/c399862d3b9d6b76c8436e924a68c45b-Abstract.html}{ImageNet
  Classification with Deep Convolutional Neural Networks}.
\newblock In {\em Advances in Neural Information Processing Systems, {NIPS}},
  pages 1106--1114, 2012.

\bibitem{kudo2018subword}
T.~Kudo.
\newblock \href{https://aclanthology.org/P18-1007/}{Subword Regularization:
  Improving Neural Network Translation Models with Multiple Subword
  Candidates}.
\newblock In I.~Gurevych and Y.~Miyao, editors, {\em {ACL}}, pages 66--75,
  2018.

\bibitem{kuehn2021ovana}
P.~Kuehn, M.~Bayer, M.~Wendelborn, and C.~Reuter.
\newblock \href{https://doi.org/10.1145/3465481.3465744}{{OVANA:} An Approach
  to Analyze and Improve the Information Quality of Vulnerability Databases}.
\newblock In {\em International Conference on Availability, Reliability and
  Security, {ARES}}, 2021.

\bibitem{BART}
M.~Lewis, Y.~Liu, N.~Goyal, M.~Ghazvininejad, A.~Mohamed, O.~Levy, V.~Stoyanov,
  and L.~Zettlemoyer.
\newblock \href{https://doi.org/10.18653/v1/2020.acl-main.703}{{BART:}
  Denoising Sequence-to-Sequence Pre-training for Natural Language Generation,
  Translation, and Comprehension}.
\newblock In {\em {ACL}}, pages 7871--7880, 2020.

\bibitem{liu2019roberta}
Y.~Liu, M.~Ott, N.~Goyal, J.~Du, M.~Joshi, D.~Chen, O.~Levy, M.~Lewis,
  L.~Zettlemoyer, and V.~Stoyanov.
\newblock \href{http://arxiv.org/abs/1907.11692}{RoBERTa: {A} Robustly
  Optimized {BERT} Pretraining Approach}.
\newblock {\em CoRR}, abs/1907.11692, 2019.

\bibitem{word2vec}
T.~Mikolov, K.~Chen, G.~Corrado, and J.~Dean.
\newblock \href{http://arxiv.org/abs/1301.3781}{Efficient Estimation of Word
  Representations in Vector Space}.
\newblock In {\em International Conference on Learning Representations,
  {ICLR}}, 2013.

\bibitem{mohaisen2013unveiling}
A.~Mohaisen and O.~Alrawi.
\newblock \href{https://doi.org/10.1145/2487788.2488056}{Unveiling Zeus:
  automated classification of malware samples}.
\newblock In {\em {WWW}}, pages 829--832, 2013.

\bibitem{mohaisen2014av}
A.~Mohaisen and O.~Alrawi.
\newblock \href{https://doi.org/10.1007/978-3-319-08509-8\_7}{AV-Meter: An
  Evaluation of Antivirus Scans and Labels}.
\newblock In {\em {DIMVA}}, pages 112--131, 2014.

\bibitem{mohaisen2015amal}
A.~Mohaisen, O.~Alrawi, and M.~Mohaisen.
\newblock \href{https://doi.org/10.1016/j.cose.2015.04.001}{{AMAL:}
  High-fidelity, behavior-based automated malware analysis and classification}.
\newblock {\em Computers \& Security.}, 52:251--266, 2015.

\bibitem{mumtaz2020learning}
S.~Mumtaz, C.~Rodr{\'{\i}}guez, B.~Benatallah, M.~Al{-}Banna, and S.~Zamanirad.
\newblock \href{https://doi.org/10.1109/IJCNN48605.2020.9207140}{Learning Word
  Representation for the Cyber Security Vulnerability Domain}.
\newblock In {\em {IJCNN}}, pages 1--8, 2020.

\bibitem{nguyen2013reliability}
V.~H. Nguyen and F.~Massacci.
\newblock \href{https://doi.org/10.1145/2484313.2484377}{The (un)reliability of
  {NVD} vulnerable versions data: an empirical experiment on Google Chrome
  vulnerabilities}.
\newblock In {\em {ACM} {ASIACCS}}, 2013.

\bibitem{niakanlahiji2018natural}
A.~Niakanlahiji, J.~Wei, and B.~Chu.
\newblock \href{https://doi.org/10.1109/BigData.2018.8622255}{A Natural
  Language Processing Based Trend Analysis of Advanced Persistent Threat
  Techniques}.
\newblock In {\em {IEEE} BigData}, pages 2995--3000, 2018.

\bibitem{GPT}
A.~Radford, K.~Narasimhan, T.~Salimans, and I.~Sutskever.
\newblock Improving language understanding by generative pre-training.
\newblock {\em OpenAI}, 2018.

\bibitem{radford2019language}
A.~Radford, J.~Wu, R.~Child, D.~Luan, D.~Amodei, I.~Sutskever, et~al.
\newblock Language models are unsupervised multitask learners.
\newblock {\em OpenAI blog}, 1(8):9, 2019.

\bibitem{T5}
C.~Raffel, N.~Shazeer, A.~Roberts, K.~Lee, S.~Narang, M.~Matena, Y.~Zhou,
  W.~Li, and P.~J. Liu.
\newblock \href{http://jmlr.org/papers/v21/20-074.html}{Exploring the Limits of
  Transfer Learning with a Unified Text-to-Text Transformer}.
\newblock {\em JMLR}, 21:140:1--140:67, 2020.

\bibitem{ref_proc15}
R.~Sennrich, B.~Haddow, and A.~Birch.
\newblock \href{https://doi.org/10.18653/v1/p16-1162}{Neural Machine
  Translation of Rare Words with Subword Units}.
\newblock In {\em {ACL}}, 2016.

\bibitem{shahid2021cvss}
M.~R. Shahid and H.~Debar.
\newblock Cvss-bert: Explainable natural language processing to determine the
  severity of a computer security vulnerability from its description.
\newblock In {\em {IEEE} ICMLA}, pages 1600--1607, 2021.

\bibitem{simonyan2014very}
K.~Simonyan and A.~Zisserman.
\newblock \href{http://arxiv.org/abs/1409.1556}{Very Deep Convolutional
  Networks for Large-Scale Image Recognition}.
\newblock In {\em International Conference on Learning Representations,
  {ICLR}}, 2015.

\bibitem{MPNet}
K.~Song, X.~Tan, T.~Qin, J.~Lu, and T.~Liu.
\newblock
  \href{https://proceedings.neurips.cc/paper/2020/hash/c3a690be93aa602ee2dc0ccab5b7b67e-Abstract.html}{MPNet:
  Masked and Permuted Pre-training for Language Understanding}.
\newblock In {\em Advances in Neural Information Processing Systems,
  {NeurIPS}}, 2020.

\bibitem{seq2seq}
I.~Sutskever, O.~Vinyals, and Q.~V. Le.
\newblock
  \href{https://proceedings.neurips.cc/paper/2014/hash/a14ac55a4f27472c5d894ec1c3c743d2-Abstract.html}{Sequence
  to Sequence Learning with Neural Networks}.
\newblock In {\em {NIPS}}, pages 3104--3112, 2014.

\bibitem{szegedy2015going}
C.~Szegedy, W.~Liu, Y.~Jia, P.~Sermanet, S.~E. Reed, D.~Anguelov, D.~Erhan,
  V.~Vanhoucke, and A.~Rabinovich.
\newblock \href{https://doi.org/10.1109/CVPR.2015.7298594}{Going deeper with
  convolutions}.
\newblock In {\em {IEEE} {CVPR}}, pages 1--9, 2015.

\bibitem{upadhyay2022mapping}
A.~Upadhyay, S.~E. Gharghasheh, and S.~Nakhodchi.
\newblock Mapping ckc model through nlp modelling for apt groups reports.
\newblock In {\em Handbook of Big Data Analytics and Forensics}, pages
  239--252. 2022.

\bibitem{transformer}
A.~Vaswani, N.~Shazeer, N.~Parmar, J.~Uszkoreit, L.~Jones, A.~N. Gomez,
  L.~Kaiser, and I.~Polosukhin.
\newblock
  \href{https://proceedings.neurips.cc/paper/2017/hash/3f5ee243547dee91fbd053c1c4a845aa-Abstract.html}{Attention
  is All you Need}.
\newblock In {\em {NeurIPS}}, pages 5998--6008, 2017.

\bibitem{wang2019learning}
Q.~Wang, B.~Li, T.~Xiao, J.~Zhu, C.~Li, D.~F. Wong, and L.~S. Chao.
\newblock \href{https://doi.org/10.18653/v1/p19-1176}{Learning Deep Transformer
  Models for Machine Translation}.
\newblock In {\em {ACL}}, pages 1810--1822, 2019.

\bibitem{ref_lncs1}
E.~W{\aa}reus and M.~Hell.
\newblock \href{https://doi.org/10.1007/978-3-030-52683-2\_1}{Automated {CPE}
  Labeling of {CVE} Summaries with Machine Learning}.
\newblock In C.~Maurice, L.~Bilge, G.~Stringhini, and N.~Neves, editors, {\em
  {DIMVA}}, pages 3--22, 2020.

\bibitem{wu2021literature}
J.~Wu.
\newblock \href{https://arxiv.org/abs/2104.11230}{Literature review on
  vulnerability detection using {NLP} technology}.
\newblock {\em CoRR}, abs/2104.11230, 2021.

\bibitem{xu2019analysis}
Y.~Xu, D.~Tran, Y.~Tian, and H.~Alemzadeh.
\newblock \href{https://doi.org/10.1109/CHASE48038.2019.00017}{Poster Abstract:
  Analysis of Cyber-Security Vulnerabilities of Interconnected Medical
  Devices}.
\newblock In {\em {IEEE/ACM} {CHASE}}, 2019.

\bibitem{XLNET}
Z.~Yang, Z.~Dai, Y.~Yang, J.~G. Carbonell, R.~Salakhutdinov, and Q.~V. Le.
\newblock
  \href{https://proceedings.neurips.cc/paper/2019/hash/dc6a7e655d7e5840e66733e9ee67cc69-Abstract.html}{XLNet:
  Generalized Autoregressive Pretraining for Language Understanding}.
\newblock In {\em {NeurIPS}}, pages 5754--5764, 2019.

\bibitem{YinTCW20}
J.~Yin, M.~Tang, J.~Cao, and H.~Wang.
\newblock Apply transfer learning to cybersecurity: Predicting exploitability
  of vulnerabilities by description.
\newblock {\em Knowl. Based Syst.}, 210:106529, 2020.

\bibitem{0003TC0YL22}
J.~Yin, M.~Tang, J.~Cao, H.~Wang, M.~You, and Y.~Lin.
\newblock Vulnerability exploitation time prediction: an integrated framework
  for dynamic imbalanced learning.
\newblock {\em World Wide Web}, 25(1):401--423, 2022.

\bibitem{YinTCYWA23}
J.~Yin, M.~Tang, J.~Cao, M.~You, H.~Wang, and M.~Alazab.
\newblock Knowledge-driven cybersecurity intelligence: Software vulnerability
  coexploitation behavior discovery.
\newblock {\em {IEEE} TII}, 19(4):5593--5601, 2023.

\bibitem{yitagesu2021automatic}
S.~Yitagesu, X.~Zhang, Z.~Feng, X.~Li, and Z.~Xing.
\newblock \href{https://doi.org/10.1109/MSR52588.2021.00016} {Automatic
  Part-of-Speech Tagging for Security Vulnerability Descriptions}.
\newblock In {\em {IEEE/ACM} {MSR}}, pages 29--40, 2021.

\bibitem{Pegasus}
J.~Zhang, Y.~Zhao, M.~Saleh, and P.~J. Liu.
\newblock \href{http://proceedings.mlr.press/v119/zhang20ae.html}{{PEGASUS:}
  Pre-training with Extracted Gap-sentences for Abstractive Summarization}.
\newblock In {\em {ICML}}, pages 11328--11339, 2020.

\end{thebibliography}
\end{document}